\documentclass[
 reprint,
 amsmath,amssymb,
 aps,
]{revtex4-2}
\usepackage{amsmath}
\usepackage{graphicx}
\usepackage{dcolumn}
\usepackage{url}
\usepackage{xcolor}
\usepackage{hyperref}

\usepackage{hyperref}
\usepackage{appendix}
\usepackage{soul}
\usepackage{xcolor}
\usepackage{amsthm}
\usepackage[ruled,linesnumbered]{algorithm2e}
\hypersetup{
    colorlinks=true,
    linkcolor=blue,
    filecolor=blue,
    urlcolor=blue,
    pdftitle={Tensor Network decoding under inter-qubit correlated errors},
    pdfauthor={Author},
}

\begin{document}

\preprint{APS/123-QED}
\title{Tensor Network decoding under inter-qubit correlated errors}

\author{Yue Yan}
\author{SiYing Wang}
\author{ZhiXin Xia}
\author{HanNuo Yuan}
\author{CanWei Shi}
\author{Xiang-Bin Wang}
\email{ xbwang@mail.tsinghua.edu.cn}
\affiliation{ State Key Laboratory of Low Dimensional Quantum Physics, Department of Physics, \\ Tsinghua University, Beijing 100084, China}

\date{\today}

\begin{abstract}

    The maximum likelihood decoder based on tensor networks has  proven highly successful for the 2D surface code, achieving the optimal decoding success rate. However, existing tensor network decoders are typically designed for independent single-qubit error models, and their performance under inter-qubit correlated error models remains unexplored. This is due to two major challenges. The first challenge lies in constructing the tensor network for correlated errors, since the same final Pauli error can arise from many different combinations of independent and correlated errors, preventing a direct factorization of the error probability. The second challenge is that even after a tensor network is constructed, it generally contains huge-dimensional tensors and is therefore not efficiently contractible. In this work, to address the first difficulty, we introduce additional binary indices and two transformations to construct a multi-index tensor network for maximum-likelihood decoding with correlated errors. To address the second difficulty, we use reparametrization, elimination, and index classification to decompose the huge-dimensional tensors into lower-dimensional tensors. This yields an efficiently contractible tensor network for error models satisfying the tractability conditions derived in this work. We perform numerical simulations for a representative correlated error model and show that the maximum-likelihood decoder implemented with our multi-index tensor network construction achieves a higher finite-size threshold than the widely used MWPM decoder.

\end{abstract}

\maketitle

\section{\label{sec:intro}introduction}

A central challenge in realizing practical quantum computers is their extreme sensitivity to noise. Currently, quantum error correction\cite{qec1,qec2,qec3,Canossa_2026} stands out as an excellent solution to the noise problem, enabling fault-tolerant processing of quantum information. This approach introduces redundant physical qubits into the quantum system to protect quantum information from noise interference. Among the various quantum error correction frameworks, the surface code\cite{surface_code_1,surface_code_2,surface_code_3} has become the mainstream foundational architecture across multiple platforms due to its high fault-tolerant threshold\cite{high_threshold_1,high_threshold_2,high_threshold_3,high_threshold_4,high_threshold_5} and ease of physical implementation\cite{ez_implementation_1,ez_implementation_2,ez_implementation_3,ez_implementation_4,ez_implementation_5,ez_implementation_6,ez_implementation_7,Wang_2024}.

As the most important indicator for the surface code, the threshold determines the major error-correction performance of the surface code. The threshold is affected by both the decoder applied and the type of noise considered. Prior works  have shown that the tensor network decoder\cite{tn_1,tn_2,tn_3,tn_4,tn_5} yields the highest threshold of the surface code under the independent single-qubit errors. Importantly, it has a higher threshold than the well-known MWPM decoder\cite{MWPM1,MWPM2,MWPM3,MWPM4} and the Union-Find decoder\cite{uf_1,uf_2,uf_3,uf_4,uf_5,uf_6} under independent single-qubit errors. However, so far it remains unknown whether the tensor network decoder has a threshold advantage under correlated errors.

Realistic quantum systems  suffer from (inter-qubit) correlated errors that arise from a variety of physical mechanisms. In particular, spatially correlated errors can be induced by non-Markovian environmental effects\cite{non-Markovian_1,non-Markovian_2}, qubit crosstalk\cite{crosstalk_1,crosstalk_2,crosstalk_3} during parallel operations, and other mechanisms. All these drastically affect the threshold. Numerical studies\cite{high_threshold_1, correlation_1,correlation_2,correlation_3} of thresholds for the MWPM decoder in the presence of correlated errors have been reported. In addition,  the maximum-likelihood threshold of the surface code under correlated errors was obtained by a statistical-mechanical mapping in Ref.\cite{correlation_3}, and the resulting threshold was recently confirmed analytically in Ref.\cite{statistical_4}. However, there are two obvious barriers in applying the tensor network decoding for correlated errors. The first challenge lies in constructing the tensor network for correlated errors, since the same final Pauli error can arise from many different combinations of independent and correlated errors, preventing a direct factorization of the error probability. The second challenge is that even after a tensor network is constructed, it generally contains huge-dimensional tensors and is therefore not efficiently contractible. Therefore, developing explicit constructions of efficiently contractible tensor networks is the central task for us to effectively apply the tensor network decoder to a surface code with correlated errors.

In this work, we present an efficiently contractible tensor network construction for maximum likelihood decoding on the 2D surface code in the presence of correlated errors. To overcome the first difficulty above, we introduce additional binary indices and two transformations to construct a multi-index tensor network. To overcome the second difficulty, we further develop a three-step procedure consisting of reparametrization, elimination, and index classification. This procedure decomposes the huge-dimensional tensors into lower-dimensional tensors, enabling efficient contraction on a classical computer. Applying the tractable tensor network constructed by our method, we perform numerical simulations for a representative correlated error model. The results show that our tensor network decoder achieves a higher finite-size threshold than the MWPM decoder.

The paper is arranged as follows: In Sec.\ref{Maximum Likelihood Decoder}, we review how the surface code performs error correction and briefly describe how the Maximum Likelihood Decoder works. In Sec.\ref{Noise model}, we introduce a fairly general correlated error model and provide a concrete example. In Sec.\ref{Sec:TN}, we explain how to explicitly construct the corresponding tensor network by introducing additional indices and two transformations. We then use three procedures, namely reparametrization, elimination and index classification, to reduce the tensor dimensions and show that the resulting tensor network can be contracted efficiently on a classical computer. We also give the conditions required for classically tractable contraction and illustrate the construction for a representative error model. In Sec.\ref{Sec:performance}, we present numerical simulations for the representative correlated error model. We further compare this threshold with the asymptotic threshold obtained from statistical methods \cite{correlation_3,statistical_4}. Finally, Sec.\ref{Sec:conclusion} concludes the paper with a summary.

\section{Maximum Likelihood Decoder in surface codes\label{Maximum Likelihood Decoder}}

The surface code is a topological quantum error correcting code that encodes logical qubits in a square lattice\cite{surface_code_1,surface_code_2,surface_code_3}. An n-qubit surface code is characterized by its stabilizer group $G$. The site stabilizers and plaquette stabilizers are defined as $\mathcal{X}_u = \prod_{i \in V_u}X_i$ and $\mathcal{Z}_p = \prod_{j \in V_p}Z_j$, where $u$ and $p$ represent the corresponding measurement qubits. Here, $V_u$ and $V_p$ denote the sets of data qubits neighboring $u$ and $p$. The stabilizer group of the surface code $G$ is generated by $\mathcal{X}_u$ stabilizers and $\mathcal{Z}_p$ stabilizers, and we denotes $G=\langle\mathcal{X}_u, \mathcal{Z}_p \rangle $. 

We use $\mathcal{P}_n$ for the group of n-qubit Pauli operators. The centralizer of $G$ is $\mathcal{C}(G)=\{c_g \in \mathcal{P}_n| [c_g, g]=0, \forall  g \in G \}$. The elements in the set $\{\mathcal{C}(G) \backslash G\}$ are logical operators of the surface code. When an error $e \in \mathcal{P}_n$ experiences in a surface code, a syndrome will occur when the error e anticommutes with a stabilizer $g \in G$. We define a  binary vector s as the syndrome, the i-th element of s satisfies $eg_i=(-1)^{s_i} g_ie, s_i \in \{0,1\}$.

We define r as the recovery operation that removes the syndrome produced by the error e, which means that $er \in \mathcal{C}(G)$. Every correctable recovery r satisfies $er = g$, others will cause a logical error in surface code.
Given a recovery r, the set of all recoveries can be classified into four mutually exclusive cosets: $\mathcal{C}(rL_I) = \{rg|g \in G\}$, $\mathcal{C}(rL_X) = \{rL_Xg|g \in G\}$, $\mathcal{C}(rL_Y) = \{rL_Yg|g \in G\}$ and $\mathcal{C}(rL_Z) = \{rL_Zg|g \in G\}$, where $L_I$ is the identity logical operator and $L_X$, $L_Y$, and $L_Z$ are the logical Pauli $X$, $Y$, $Z$ operators.
A Maximum Likelihood Decoder selects the coset with the highest probability and applies any recovery from that coset\cite{tn_1}. Consequently, for a given syndrome s, the conditional logical failure probability $p_{fail}(s)$ is 

\begin{equation}
    p_{fail}(s) = 1 - \frac{\max_{j \in \{I,X,Y,Z\}}p(\mathcal{C}(rL_j))}{\sum_{j \in \{I,X,Y,Z\}}p(\mathcal{C}(rL_j))} 
\end{equation}

where r is a recovery associated with the syndrome s, $p(\mathcal{C}(rL_j))$ represents the total probability of all recoveries in cosets $\mathcal{C}(rL_j)$. In the following, we construct tensor networks to compute $p(\mathcal{C}(rL_j))$ under correlated noise. The overall logical error rate is then estimated numerically by sampling errors.

\section{Correlated error Noise model \label{Noise model}}

Consider a general correlated noise model consisting of multiple independent noise channels.

The first channel is an independent single-qubit Pauli noise channel, where each data qubit independently experiences a Pauli error with probability $p_1$. Let $p_{1x},p_{1y}$ and $p_{1z}$ denote the probabilities of Pauli X,Y and Z errors, respectively. The total error probability is defined as $p_1 = p_{1x}+p_{1y}+p_{1z}$. The channel is

\begin{equation}
    \mathcal{E}_1(\rho) = (1-p_1)\rho + p_{1x}X\rho X + p_{1y}Y\rho Y + p_{1z}Z\rho Z
\end{equation}

 and more generally, the i-th noise channel is

\begin{equation}
    \mathcal{E}_i(\rho ) = (1-p_i)\rho + \sum_j p_{ij} E_{ij}\rho E_{ij}^\dagger
    \label{channel_e}
\end{equation}
where $E_{ij}$ is the correlated error operator for which we would go into more details soon, $p_{ij}$ denote the probability of the error $E_{ij}$, and $p_i = \sum_j p_{ij}$. 

For a fixed i, each correlated error operator $E_{ij} \in \mathcal{P}_n$ acts on the same set of $m_i$ data qubits. We denote this set by $Q_i = \{i_1^q,i_2^q,...,i_{m_i}^q\}$, which is independent of $j$.
Ignoring the identity operator, the error $E_{ij}$ can be expressed as

\begin{equation}
    E_{ij} = e_{ij}(i_1^q)e_{ij}(i_2^q)...e_{ij}(i_{m_i}^q)
    \label{eq_eij}
\end{equation}

The overall noise model is given by their composition $\mathcal{E} = \mathcal{E}_1 \circ \mathcal{E}_2 \circ ... \circ \mathcal{E}_l$, meaning that each data qubit can be affected by both independent and correlated errors simultaneously.

\begin{figure}
    \centering
    \includegraphics[width=1.0\linewidth]{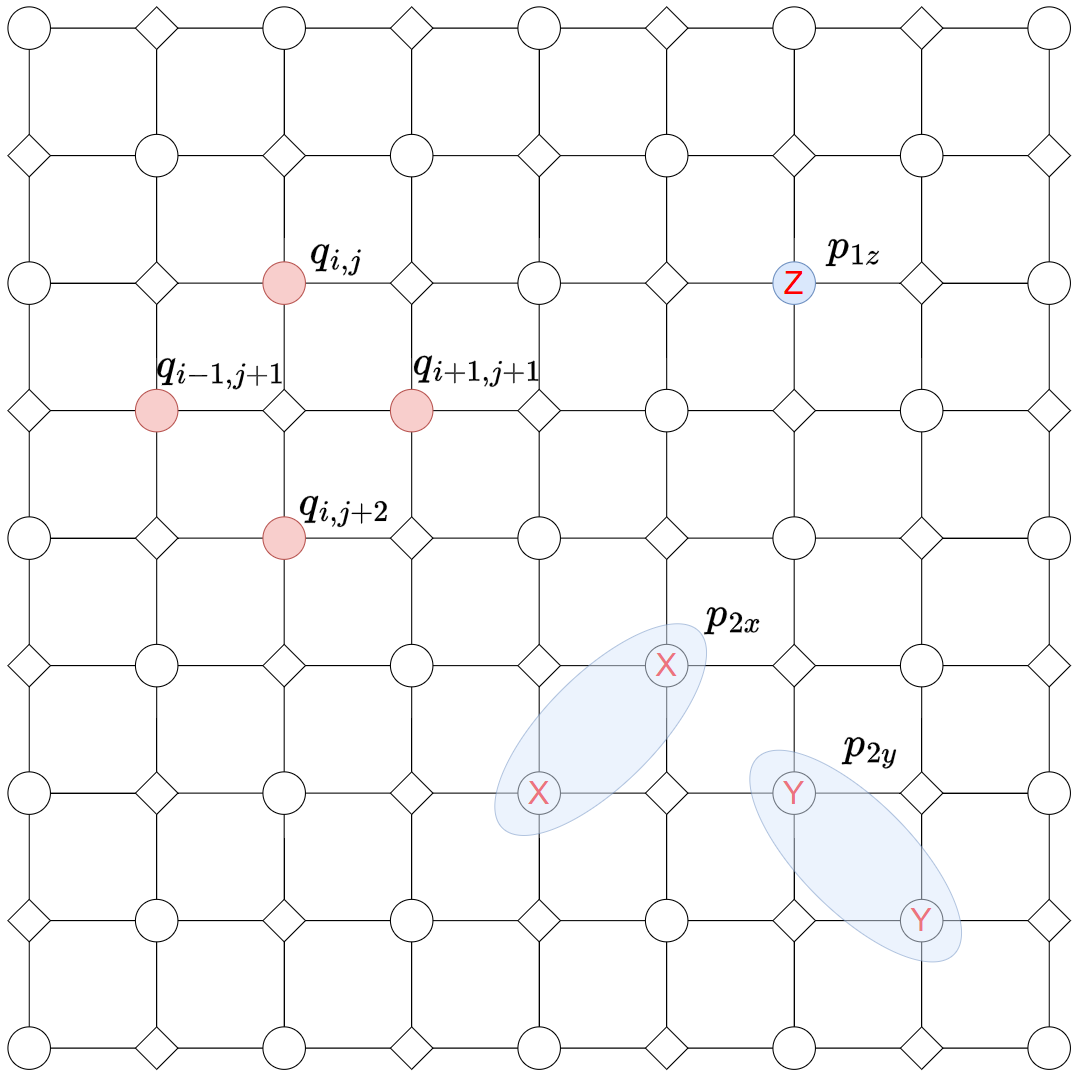}
    \caption{An example of error model ENN. Diamonds represent measurement qubits, and circles represent data qubits. The four pink circles correspond to $q_{i,j}, q_{i-1,j+1}, q_{i,j+2}, q_{i+1,j+1}$, where the subscripts denote the coordinates of the qubits. For example, $\{q_{i+1,j+1}, q_{i,j+2}\}$ is a pair of nearest-neighbor data qubits. The blue circle in the upper-right part indicates that an independent single-qubit Pauli Z error occurs at that data qubit, with probability $p_{1z}$. The two pairs of nearest-neighbor data qubits enclosed by the blue ellipse in the lower-right part experience a correlated XX error and correlated YY error, with probabilities $p_{2x}$ and $p_{2y}$, respectively.}
    \label{fig:noise model}
\end{figure}

As an example of the formalism above, we consider a correlated error model between nearest-neighbor data qubits. For convenience, we name it error model ENN. The first channel $\mathcal{E}_1$ is also an independent single-qubit  Pauli noise model. The second channel $\mathcal{E}_2$ consists of correlated errors between nearest-neighbor data qubits, as shown in Eq.\ref{e2} illustrated in Fig\ref{fig:noise model}. A correlated Pauli error occurs independently with probability $p_2$ on each of the nearest-neighbor data-qubit pairs $\{q_{i,j}, q_{i-1,j+1}\}$ and $\{q_{i,j}, q_{i+1,j+1}\}$. Similarly, $p_{2x},p_{2y}$ and $p_{2z}$ denote the probabilities of correlated Pauli XX,YY and ZZ errors, respectively. And the probability $p_2 = p_{2x}+p_{2y}+p_{2z}$. The channel is

\begin{align}
    \mathcal{E}_2(\rho ) =& (1-p_2)\rho + p_{2x}X_{i}X_{j}\rho X_{i}X_{j} + p_{2y}Y_{i}Y_{j}\rho Y_{i}Y_{j} \nonumber \\
    &+ p_{2z}Z_{i}Z_{j}\rho Z_{i}Z_{j}
    \label{e2}
\end{align}

In the ENN error model, independent single-qubit Pauli errors and nearest-neighbor correlated Pauli errors are both present. Each data qubit is affected by the noise channel $\mathcal{E}_1$, while each nearest-neighbor data qubit pair is independently affected by a correlated two-qubit noise channel of the form in Eq.\ref{e2}.

\section{Optimal Decoding with Multi-Index Tensor Networks\label{Sec:TN}}

\subsection{Global Tensor Network Construction\label{Sec:global_TN}}

Under the independent single-qubit Pauli noise model, the probabilities of the four cosets $\mathcal{C}(rL_I)$, $\mathcal{C}(rL_X)$, $\mathcal{C}(rL_Y)$, $\mathcal{C}(rL_Z)$ in Sec.\ref{Maximum Likelihood Decoder} can be represented by tensor networks\cite{tn_1}. In this case, straightforwardly,   the probability of a multi-qubit Pauli error in a coset can be factorized into a product of independent single-qubit error probabilities. And hence, the tensor network can be constructed directly for decoding, as is well known. However, in the presence of correlated errors, there is no straightforward way to factorize the probability into  a product of  independent single-qubit error probabilities and  correlated qubit error probabilities. Moreover, when independent single-qubit errors and correlated errors coexist, the same final Pauli error may arise from many different combinations of these errors. Therefore, construction of tensor network for decoding in the presence of correlated errors is nontrivial, and so far it is known yet how to make it.  

To address this difficulty, we introduce additional binary indices and two transformations shown in Eq.\ref{trans_1}, Eq.\ref{trans_2} to construct the multi-index tensor network for the noise channel $\mathcal{E}$. Details are shown below.

\subsubsection{preparation of notations}

Let $r'$ denote any one of $rL_I$, $rL_X$, $rL_Y$, $rL_Z$. The probability of the cosets $\mathcal{C}(rL_I)$, $\mathcal{C}(rL_X)$, $\mathcal{C}(rL_Y)$, $\mathcal{C}(rL_X)$ defined in Sec.\ref{Maximum Likelihood Decoder} are:

\begin{equation}
    p(\mathcal{C}(r')) = \sum_{g \in G} p(\prod_{k} r'_k g_k)
    \label{eq_gk}
\end{equation}

where $r'_k$ and $g_k \in G$ denote the Pauli operators acting on the the k-th data qubit.

Any stabilizer $g \in G$ can be uniquely determined by two binary vectors $x,z$:
\begin{equation}
    g(x, z) = \prod_u (\mathcal{X}_u)^{x_u} \cdot \prod_p (\mathcal{Z}_p)^{z_p}
\end{equation}

where $x_u,z_p \in \{0,1\}$.

Each data qubit is connected to four measurement qubits, two of which correspond to $\mathcal{X}_u$ stabilizers and the other two correspond to $\mathcal{Z}_p$ stabilizers. The operator $g_k$ in Eq.\ref{eq_gk} can be uniquely specified by the four binary indices associated with these neighboring measurement qubits. Let $u(k), u'(k)$ index the two $\mathcal{X}_u$ stabilizers connected to the k-th data qubit, and $p(k), p'(k)$ index the two $\mathcal{Z}_p$ stabilizers. The corresponding binary indices are $x_{u(k)}, x_{u'(k)}, z_{p(k)}, z_{p'(k)}$. Then

\begin{equation}
    g_k(x,z) = g_k(x_{u(k)}, x_{u'(k)}, z_{p(k)}, z_{p'(k)})
\end{equation}

The indices are subject to the consistency constraints
\begin{equation}
    x_{u(k)} = x_{u(k')}, \quad z_{p(l)} = z_{p(l')}
    \label{condition0}
\end{equation}
for any two data qubits $k$ and $k'$ that correspond to the same $\mathcal{X}_u$ stabilizer or any two data qubits $l$ and $l'$ that correspond to the same $\mathcal{Z}_p$ stabilizer.

On the  boundaries, the situation is similar. Only three of the indices are required to determine the operator, while at the corners, only two indices suffice.

Moreover, we use the former four indices $x_{u(k)}$, $x_{u'(k)}$, $z_{p(k)}$, $z_{p'(k)}$ and the Pauli operator $r_k'$ to uniquely determine the specific Pauli operator $O_k = r'_k g_k$ that acts on the data qubit $k$. 
Operator $O_k$ satisfies:

\begin{equation}
    O_k = r_k' \cdot X^{x_{u(k)} \oplus x_{u'(k)}}Z^{z_{p(k)} \oplus z_{p'(k)}}.
    \label{condition2_origin}
\end{equation}

To accurately compute $p(\prod_k r'_k g_k) = p(\prod_k O_k)$, We will introduce additional indices and reformulate the expression for the correlated error probability. We define a set of binary indices $D = \{d_{ij}\}$, and $d_{ij}$ takes the value 1 if the error operator $E_{ij}$ is selected in the i-th noise channel, and 0 otherwise. For convenience, when $D$ is fixed, we use $d_{ij}$ instead of $d_{ij}(D)$. All sums and products involving $d_{ij}$ are understood to be taken over the components of the same $D$.
Since at most one error operator can be selected from the same noise channel, these indices satisfy

\begin{equation}
    \sum_j d_{ij} \in \{0,1\}, d_{ij} \in \{0, 1\}.
    \label{condition1}
\end{equation}

We define the set $\tilde{D} $ such that, for any $D$, if every element of $D$ satisfies Eq.\ref{condition1}, then $D\in \tilde{D}$. This means that $\tilde{D}$ is a set that contains all elements($D$) which can represent operator $G_D$. The channel $\mathcal{E}$ can be rewritten as:

\begin{equation}
    \mathcal{E}(\rho) = \sum_{D \in \tilde{D}} p_D G_{D} \rho G_{D}^{\dagger }
    \label{channel}
\end{equation}
where 
\begin{equation}
    G_{D} = \prod_{ij}E_{ij}^{d_{ij}}
\end{equation}
and 
\begin{equation}
    p_D=\prod_i[(1-p_i)^{1-\sum_j d_{ij}}\prod_j p_{ij}^{d_{ij}}].
    \label{eq_pok}
\end{equation}
A detailed derivation of Eq.~\ref{channel} is provided in the Appendix.

Recall that our goal is to compute $p(\mathcal{C}(r'))$. According to Eq.\ref{eq_gk}, this amounts to summing the probabilities $p(\prod_k O_k)$ over all Pauli operator strings $\{\prod_k O_k\}$ in the coset $\mathcal{C}(r')$. For a Pauli operator $\prod_k O_k$, its probability $p(\prod_k O_k)$ contributes to $p(\mathcal{C}(r'))$ only when $\prod_k O_k$ equal to a certain operator $G_D$, that is $G_D=\prod_k O_k$.
Using Eq.\ref{eq_eij} and Eq.\ref{condition2_origin}, for any k, this requirement leads to:

\begin{equation}
    r_k' \cdot X^{x_{u(k)} \oplus x_{u'(k)}}Z^{z_{p(k)} \oplus z_{p'(k)}} = \prod_{ij} {e_{ij}(k)}^{d_{ij}}
    \label{condition2}
\end{equation}

Note that elements of any D satisfy Eq.\ref{condition1}, we define the set $E_{valid}$ such that, for any $D$, if elements of $D$ satisfy Eq.\ref{condition2}, then $D\in E_{valid}$.
We define $p(\prod_{k} O_k) = \sum_{D \in E_{valid}} p^*(\prod_{k} O_k)$ and under this definition $p^*(\prod_{k} O_k) = p_D$. 

\subsubsection{details of the global tensor networks construction}

We introduce two transformations to rewrite $p_{ij}$ and $(1-p_i)$ in Eq.(\ref{channel_e}) as a product of $m_i$ identical factors:

\begin{equation}
    p_{ij} = \prod_{k \in Q_i} \alpha_{ij}(k) 
    \label{trans_1}
\end{equation}

\begin{equation}
    (1-p_i) = \prod_{k \in Q_i} \beta_i(k)
    \label{trans_2}
\end{equation}
where $\alpha_{ij}(k)= {p_{ij}}^{\frac{1}{m_i}}$ and $\beta_i(k)  = {(1-p_i)}^{\frac{1}{m_i}}$ for $k \in Q_i$. Note that k is independent of j.

For every $k \in Q_i$, we introduce a new index $d_{ijk}$ defined by $d_{ijk} = d_{ij}$. This means that these indices must satisfy the conditions:

\begin{equation}
    d_{ijs} = d_{ijt}(s,t \in Q_i)
    \label{ijk_condition0}
\end{equation}

\begin{equation}
    \sum_j d_{ijk} \in \{0,1\}, d_{ijk} \in \{0, 1\}
    \label{ijk_condition1}
\end{equation}

\begin{equation}
    r_k' \cdot X^{x_{u(k)} \oplus x_{u'(k)}}Z^{z_{p(k)} \oplus z_{p'(k)}} = \prod_{ij} {e_{ij}(k)}^{d_{ijk}}
    \label{ijk_condition2}
\end{equation}

We define a set of binary indices $D' = \{d_{ijk}\}$ and the set $\tilde{E}_{valid}$ such that, for any $D'$, if  elements of $D'$ satisfy Eq.\ref{ijk_condition0}, Eq.\ref{ijk_condition1} and Eq.\ref{ijk_condition2}, then $D'\in \tilde{E}_{valid}$.

Substituting the above two transformations into Eq.\ref{eq_pok}, $p^*(\prod_k O_k)$ can be rewritten as: 
\begin{eqnarray}
    p^*(\prod_k O_k) &=   \prod_k \prod_{i} [\prod_{j} ({p_{ij}}^{\frac{1}{m_i}})^{d_{ijk}}] \nonumber \\
    & ({(1-p_i)}^{\frac{1}{m_i}})^{1-\sum_j d_{ijk}} 
    \label{eq_pok2}
\end{eqnarray}

Eq.\ref{eq_pok2} can be rewritten as  $p^*(\prod_k O_k) = \prod_k p^*(O_k)$ where $p^*(O_k) = \prod_{i} [\prod_{j} ({p_{ij}}^{\frac{1}{m_i}})^{d_{ijk}}]({(1-p_i)}^{\frac{1}{m_i}})^{1-\sum_j d_{ijk}}$.

Now, Eq.\ref{eq_gk} can be expressed in terms of the relevant indices $x_{u(k)}$, $x_{u'(k)}$, $z_{p(k)}$, $z_{p'(k)}$, $d_{ijk}$ and operator $r_k'$ as follows:
\begin{align}
    p(\mathcal{C}(r')) &= \sum_{g \in G} \sum_{D' \in \tilde{E}_{valid}} \prod_{k}  \nonumber \\
    &\mathcal{F}_{x_{u(k)}, x_{u'(k)}, z_{p(k)}, z_{p'(k)}, d_{ijk}}(r_k')
\end{align}
where
$\mathcal{F}_{x_{u(k)}, x_{u'(k)}, z_{p(k)}, z_{p'(k)}, d_{ijk}}(r_k') = p^*(O_k)$.

We define $\Lambda=\{x_{u(k)},x_{u'(k)},z_{p(k)},z_{p'(k)},d_{ijk}\}_{i,j,k}$, which is the multi-index set that contains all indices appearing in the conditions(Eq.\ref{condition0}, Eq.\ref{ijk_condition0}, Eq.\ref{ijk_condition1}, Eq.\ref{ijk_condition2}). We introduce a global auxiliary tensor $T_{\Lambda}$ which takes the value 1 when the indices satisfy all  conditions of Eq.\ref{condition0}, Eq.\ref{ijk_condition0}, Eq.\ref{ijk_condition1} and Eq.\ref{ijk_condition2}, and 0 otherwise. With this auxiliary tensor constructed, $p(\mathcal{C}(r'))$ can be written as a summation over tensor indices below:

\begin{align}
    p(\mathcal{C}(r')) =& \sum_{\Lambda} T_{\Lambda} \prod_{k} \nonumber \\
    & \mathcal{F}_{x_{u(k)}, x_{u'(k)}, z_{p(k)}, z_{p'(k)}, d_{ijk}}(r_k').
    \label{sum_tn}
\end{align}

\begin{figure}
    \centering
    \includegraphics[width=0.9\linewidth]{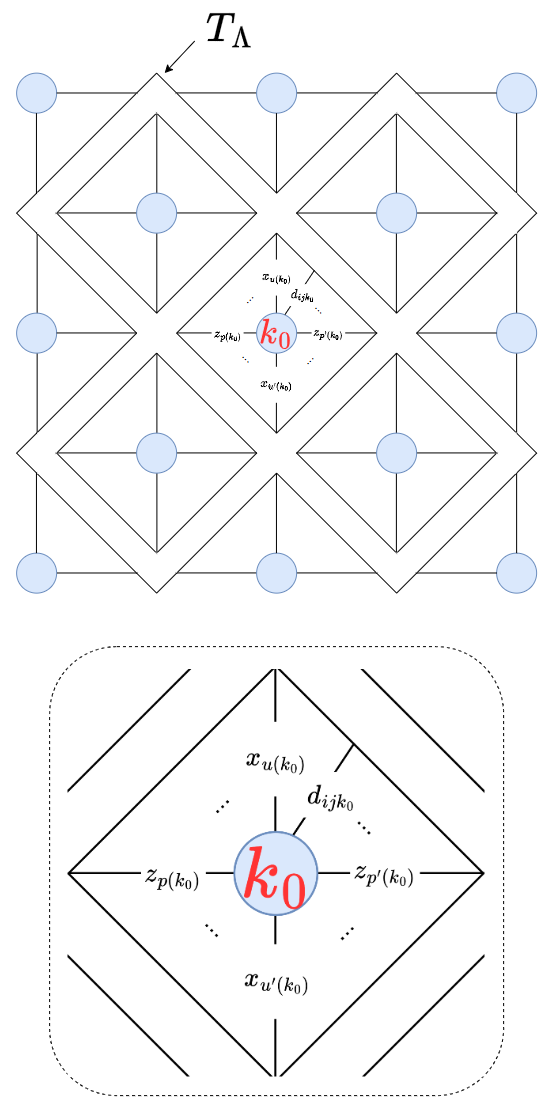}
    \caption{The tensor network corresponding to the specific surface code with $d=3$ under a general correlated error model. The dashed box shows a magnified view of the central region of the tensor network. The central blue circle  represents the tensor $\mathcal{F}_{x_{u(k_0)}, x_{u'(k_0)}, z_{p(k_0)}, z_{p'(k_0)}, d_{ijk_0}}(r_{k_0}')$, and the surrounding box represents a tensor $T_{\Lambda}$ of huge dimension. In the figure, only the indices $d_{ijk}$ with $k = k_0$  corresponding to the specific $k_0$-th data qubit are labeled, the other $d_{ijk}$ with $k \neq k_0$ indices are omitted.}
    \label{fig:big tn}
\end{figure}

This summation corresponds to the tensor network shown in Fig.\ref{fig:big tn}. By contracting the tensor network, we can compute $p(\mathcal{C}(r'))$ and perform exact decoding in principle.

\subsection{Constructing an Efficiently Contractible Tensor Network\label{Sec:CTN}}

The tensor $T_{\Lambda}$ constructed so far cannot be contracted within a reasonable computing time using a classical computer. The key reason is that $T_{\Lambda}$ has a huge number of indices, on the order of $(\sum_i m_i)n$. Each index is binary, meaning that the overall dimension of the auxiliary tensor $T_{\Lambda}$ is $O(2^{(\sum_i m_i)n})$, which is computationally prohibitive. In the following, we present a method to solve this key problem.

{\em The main idea}. There are three major steps in our construction of an efficient Contractible tensor network.

1. Reparametrization: We shall first reparametrize the indices $\{d_{ijk}\}$.

2. Elimination: We eliminate the indices $x_{u(k)}$, $x_{u'(k)}$, $z_{p(k)}$, $z_{p'(k)}$. 

3. Index Classification: We classify all indices by a graph and construct tensors of tractable dimensions. We also present conditions on the error model $\mathcal{E}$ under which the corresponding tensor network can be decomposed into smaller-dimensional tensors and thus contracted efficiently on a classical computer. 

An example of the error model ENN is given to improve readability.

{\em Reparametrization.} We introduce two binary indices $r_x^{idx}, r_z^{idx}$ and  define the Pauli operator $r_k'$  as $r_k' = X^{r_x^{idx}}Z^{r_z^{idx}}$. The operator $O_k$ in Eq.\ref{condition2_origin} can be expanded as
\begin{equation}
    O_k = X^{r_x^{idx} \oplus x_{u(k)} \oplus x_{u'(k)}}Z^{r_z^{idx} \oplus z_{p(k)} \oplus z_{p'(k)}}.
\end{equation}

{\em Reparametrization} $\mathcal{P}1$: With three indices($d_{1xk}$, $d_{1yk}$, $d_{1zk}$), we define $O_{1k} = X^{d_{1xk}} Y^{d_{1yk}} Z^{d_{1zk}}$ as the error operator induced by an independent single-qubit error. We introduce reparametrization 
$d_{1xk}$, $d_{1yk}$, $d_{1zk}$ $\{x_{u(k)}^*, x_{u'(k)}^*, z_{p(k)}^*, z_{p'(k)}^*\}$. 
For convenience, the operator $O_{1k}$ is written as
\begin{equation}
    O_{1k} = X^{r_x^{idx} \oplus x_{u(k)}^* \oplus  x_{u'(k)}^*} Z^{r_z^{idx} \oplus z_{p(k)}^* \oplus  z_{p'(k)}^*}.
\end{equation}

However, without any restrictions on $\{x_{u(k)}^*, x_{u'(k)}^*, z_{p(k)}^*, z_{p'(k)}^*\}$, the reparametrization by them is redundant, because the map from $\{x_{u(k)}^*, x_{u'(k)}^*, z_{p(k)}^*, z_{p'(k)}^*\}$ to $\{d_{1xk}, d_{1yk}, d_{1zk} \}$ is many-to-one rather than one-to-one.

Consider the fixed values of $d_{1xk}$, $d_{1yk}$, $d_{1zk}$ and an assignment of $A_{k1} = \{x_{u(k)}^*, x_{u'(k)}^*, z_{p(k)}^*, z_{p'(k)}^*\}$ that satisfy

\begin{equation}
    X^{d_{1xk}} Y^{d_{1yk}} Z^{d_{1zk}} =  X^{r_x^{idx} \oplus x_{u(k)}^* \oplus  x_{u'(k)}^*} Z^{r_z^{idx} \oplus z_{p(k)}^* \oplus  z_{p'(k)}^*}
    \label{sub_4idx}
\end{equation}
where overall phase factors $\pm 1$ are ignored. Then, the assignments of $A_{k2}=\{1-x_{u(k)}^*, 1-x_{u'(k)}^*, z_{p(k)}^*, z_{p'(k)}^*\}$, $A_{k3}=\{x_{u(k)}^*, x_{u'(k)}^*, 1-z_{p(k)}^*, 1-z_{p'(k)}^*\}$ and $A_{k4}=\{1-x_{u(k)}^*, 1-x_{u'(k)}^*, 1-z_{p(k)}^*, 1-z_{p'(k)}^*\}$ corresponding to the same operator $O_{1k} = X^{d_{1xk}} Y^{d_{1yk}} Z^{d_{1zk}}$ according to Eq.\ref{sub_4idx}. Therefore, when $d_{1xk}$, $d_{1yk}$, $d_{1zk}$ are reparametrized by the set of binary indices $\{x_{u(k)}^*, x_{u'(k)}^*, z_{p(k)}^*, z_{p'(k)}^*\}$, only one assignment in $\{A_{k1}, A_{k2}, A_{k3}, A_{k4}\}$ is needed in calculation of the tensor network, since the summation over $A_{kt}$ is the same for $t=1,2,3,4$ in the contraction calculation according to Eq.\ref{sum_tn}.

Under the reparametrization $\mathcal{P}1$, the probability $p^*(O_k)$  below Eq.\ref{eq_pok2} can be reformulated by $p^*(O_{ik})$:
\begin{equation}
    p^*(O_k) = \prod_i p^*(O_{ik})
\end{equation}

where 

\begin{equation}
    p^*(O_{ik}) = [\prod_{j} ({p_{ij}}^{\frac{1}{m_i}})^{d_{ijk}}] ({(1-p_i)}^{\frac{1}{m_i}})^{1-\sum_j d_{ijk}}
    \label{p_oik}
\end{equation}

For $i=1$, $p^*(O_{1k})$ is modified as:
\begin{eqnarray}
    p^*(O_{1k}) =& (1-\overline{x})(1-\overline{z})(1-p_1) + \overline{x}(1-\overline{z})p_{1x} +  \nonumber \\ 
    &(1-\overline{x})\overline{z}p_{1z} + \overline{x}\cdot\overline{z}p_{1y}
    \label{p_o1k}
\end{eqnarray}

where 

\begin{align}
    \overline{x} &=  r_x^{idx} \oplus x_{u(k)}^* \oplus  x_{u'(k)}^* \nonumber \\
    \overline{z} &= r_z^{idx} \oplus z_{p(k)}^* \oplus  z_{p'(k)}^*.
\end{align}

{\em Reparametrization} $\mathcal{P}2$:  For the indices corresponding to correlated errors($i>1$), we use this reparametrization.  The indices $d_{ijk}$ corresponding to $e_{ij}(k) = X$ or $e_{ij}(k) = Z$ remain unchanged,  while the index $d_{ijk}$ corresponding to $e_{ij}(k) = Y$ is decomposed into two indices $d_{ijk,yx}$ and $d_{ijk,yz}$. Obviously, we have
\begin{equation}
    d_{ijk,yx} = d_{ijk,yz} = d_{ijk} (e_{ij}(k) = Y)
    \label{add_condition1}
\end{equation}
After performing the reparametrization $\mathcal{P}1$ and $\mathcal{P}2$ for $d_{ijk}$, Eq.\ref{ijk_condition2} can be rewritten as:

\begin{equation}
    \left\{
        \begin{aligned}
        x_{u(k)} \oplus x_{u'(k)} &= x_{u(k)}^* \oplus  x_{u'(k)}^* \oplus [(\sum_{ij} f^x_{ijk}) \quad mod 2]\\
        z_{p(k)} \oplus z_{p'(k)} &= z_{p(k)}^* \oplus  z_{p'(k)}^* \oplus [(\sum_{ij} f^z_{ijk}) \quad mod 2]
        \end{aligned}
    \right.
    \label{re_condition2}
\end{equation}
where
\begin{equation}
    f^x_{ijk} = \left\{
        \begin{aligned}
            d_{ijk} \quad (e_{ij}(k) = X)\\
            d_{ijk,yx} \quad (e_{ij}(k) = Y)
        \end{aligned}
    \right.
\end{equation}
and 
\begin{equation}
    f^z_{ijk} = \left\{
        \begin{aligned}
            d_{ijk} \quad (e_{ij}(k) = Z)\\
            d_{ijk,yz} \quad (e_{ij}(k) = Y)
        \end{aligned}.
    \right.
\end{equation}

We can divide the set of indices $\{f^x_{ijk}\}$ into two disjoint subsets: $\{f^{x1}_{ijk}\}$ and $\{f^{x2}_{ijk}\}$. Therefore, the upper equation in Eq.\ref{re_condition2} can be written as:
\begin{align}
    x_{u(k)} \oplus x_{u'(k)} =& \{ x_{u(k)}^* \oplus [(\sum_{ij} f^{x1}_{ijk}) \quad mod \quad 2] \} \oplus \nonumber \\
    & \{ x_{u'(k)}^* \oplus [(\sum_{ij} f^{x2}_{ijk}) \quad mod \quad 2] \}
\end{align}

{\em Elimination.} As mentioned in reparametrization $\mathcal{P}1$, only one assignment in $\{A_{k1}, A_{k2}, A_{k3}, A_{k4}\}$ is needed in calculation of the tensor network. Note that there is exactly one assignment that makes the following condition hold:
\begin{equation}
    \left\{
        \begin{aligned}
        x_{u(k)}  &= x_{u(k)}^* \oplus [(\sum_{ij} f^{x1}_{ijk}) \quad mod \quad 2] \\
        x_{u'(k)} &= x_{u'(k)}^* \oplus [(\sum_{ij} f^{x2}_{ijk}) \quad mod \quad 2]
        \end{aligned}
    \right.
    \label{substitution1}
\end{equation}
\begin{equation}
    \left\{
        \begin{aligned}
        z_{p(k)}  &= z_{p(k)}^* \oplus [(\sum_{ij} f^{z1}_{ijk}) \quad mod \quad 2] \\
        z_{p'(k)} &= z_{p'(k)}^* \oplus [(\sum_{ij} f^{z2}_{ijk}) \quad mod \quad 2]
        \end{aligned}
    \right.
    \label{substitution2}
\end{equation}
where $\{f^{z1}_{ijk}\}$ and $\{f^{z2}_{ijk}\}$ are disjoint subsets of $\{f^{z}_{ijk}\}$.

\begin{figure}
    \centering
    \includegraphics[width=1.0\linewidth]{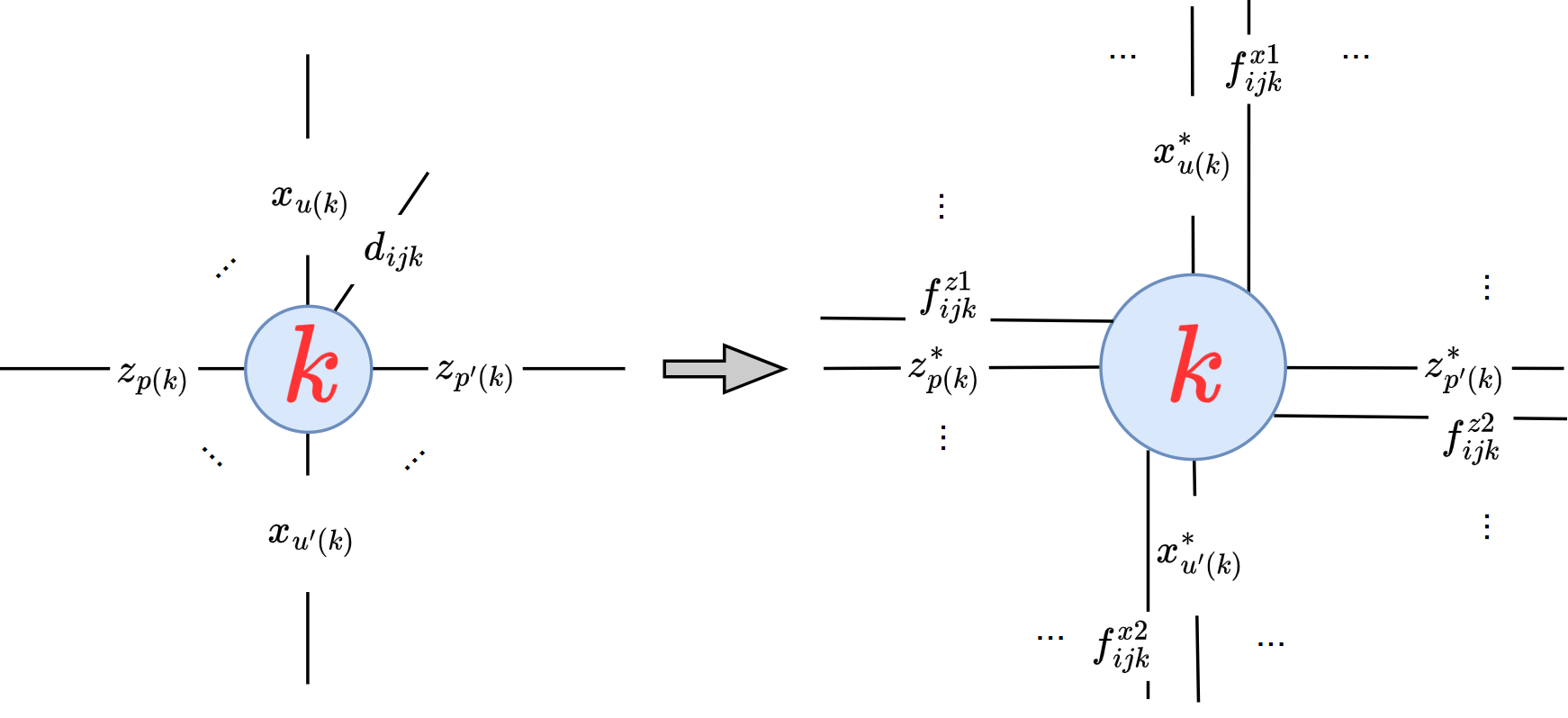}
    \caption{Under reparametrization $\mathcal{P}1$, $\mathcal{P}2$ and elimination, the right tensor $\mathcal{F}_{x_{u(k)}, x_{u'(k)}, z_{p(k)}, z_{p'(k)}, d_{ijk}}(r_k')$ in Eq.\ref{sum_tn} is replaced by the left tensor $\mathcal{F}_{x_{u(k)}^*,x_{u'(k)}^*,z_{p(k)}^*,z_{p'(k)}^*,f_{ijk}^{x1}, f_{ijk}^{x2},f_{ijk}^{z1}, f_{ijk}^{z2}}(r_k')$.}
    \label{fig:change_tn}
\end{figure}

According to Eq.\ref{substitution1} and Eq.\ref{substitution2}, we can choose to use the indices at the right hand side of these equations instead of indices $x_{u(k)}$, $x_{u'(k)}$, $z_{p(k)}$, $z_{p'(k)}$ in Eq.\ref{condition0}.
Under this choice, condition Eq.\ref{ijk_condition2} is automatically satisfied, and the tensor $\mathcal{F}_{x_{u(k)}, x_{u'(k)}, z_{p(k)}, z_{p'(k)}, d_{ijk}}(r_k')$ in Eq.\ref{sum_tn} is replaced by $\mathcal{F}_{x_{u(k)}^*,x_{u'(k)}^*,z_{p(k)}^*,z_{p'(k)}^*,f_{ijk}^{x1}, f_{ijk}^{x2},f_{ijk}^{z1}, f_{ijk}^{z2}}(r_k')$, as shown in Fig.\ref{fig:change_tn}.

{\em  Index Classification.} Define a multi-index set $\Lambda' = \{x_{u(k)}^*,x_{u'(k)}^*,z_{p(k)}^*,z_{p'(k)}^*,f_{ijk}^{x1}, f_{ijk}^{x2}, f_{ijk}^{z1}, f_{ijk}^{z2}\}_{i,j,k}$. We classify the indices by constructing a graph $\Gamma=(V_\Gamma,L_\Gamma)$: each index in $\Lambda'$ is represented by a vertex $V_\Gamma$, and for any two vertices $u,v \in V_\Gamma$, we add an edge $l \in L_\Gamma$ between them if the corresponding indices appear in the same condition. The connected components of $\Gamma$ then provide a classification of the indices: the indices corresponding to the same connected component are in the same class. For every connected component $\Gamma^*=(V_\Gamma^*,L_\Gamma^*)$, we define a multi-index set $\Lambda^*$, whose elements are the indices corresponding to the vertices in $V_\Gamma^*$. 

Because different connected components correspond to different conditions, for convenience, we define a condition set $S$ containing all the conditions(Eq.\ref{condition0}, Eq.\ref{ijk_condition0}, Eq.\ref{ijk_condition1}, Eq.\ref{add_condition1}, Eq.\ref{substitution1}, Eq.\ref{substitution2}). For every $\Gamma^*$, we define $S^* \subset S$ as the subset containing the conditions involving the indices in $\Lambda^*$. Similar to $T_{\Lambda}$, for every connected component $\Gamma^*$, we construct an auxiliary tensor $T_{\Lambda^*}$ which takes the value 1 when the conditions in $S^*$ are satisfied, and 0 otherwise. The dimension of the tensor $T_{\Lambda^*}$ is $2^{n_V^*}$ , where $n_V^*$ is the number of vertices in $\Gamma^*$. Given these, we can replace the original high-dimensional tensor $T_{\Lambda}$ by a set of lower-dimensional tensors, which can significantly reduce the computational cost. In conclusion, for a given error model $\mathcal{E}$, after reparametrization, elimination and index classification, if every $n_V^*$ of $\Gamma^*$ is a constant, we can construct the tensor network that can be contracted efficiently by using the method of Ref.\cite{tn_2}.

\begin{figure}
    \centering
    \includegraphics[width=1.0\linewidth]{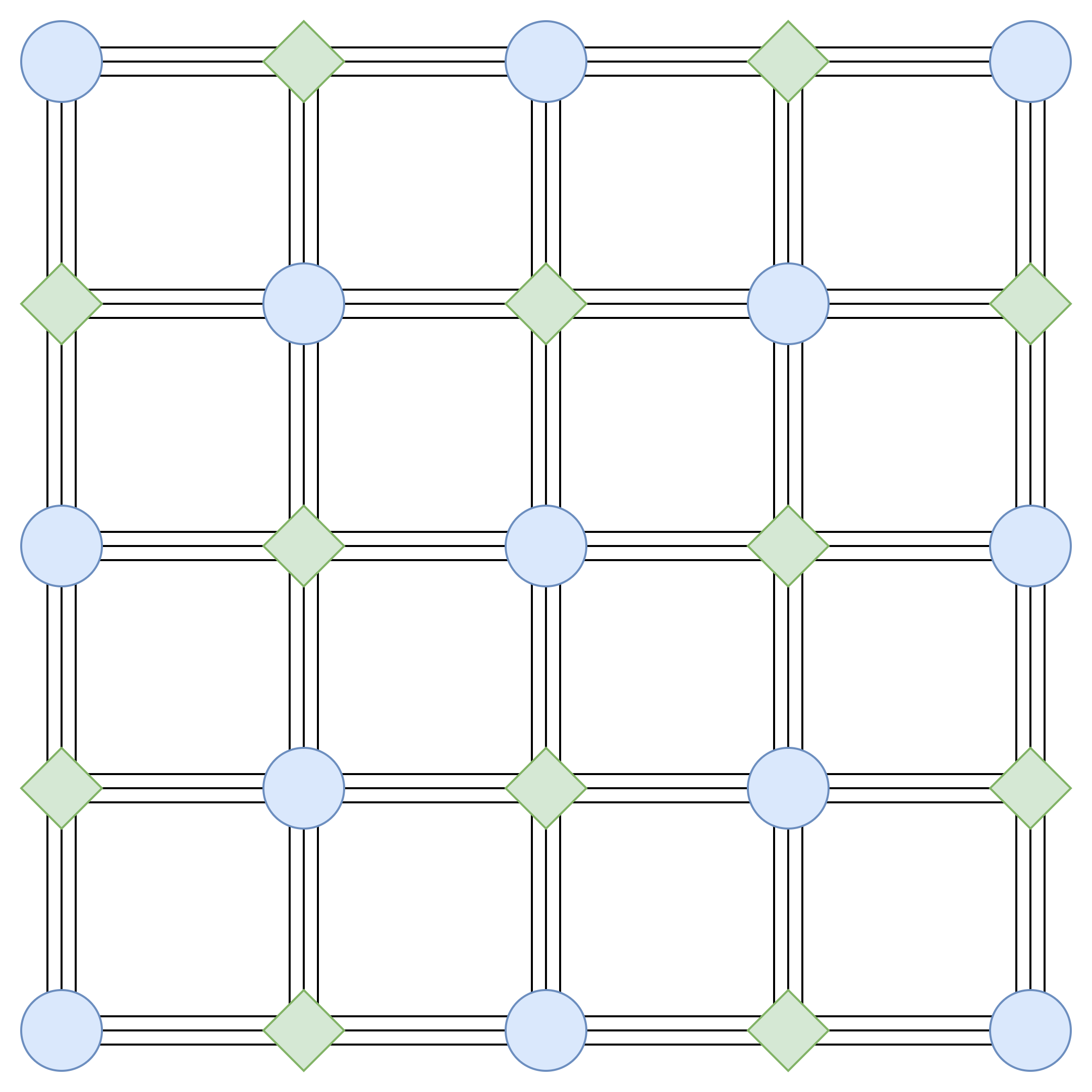}
    \caption{The tensor network corresponding to the surface code with $d=3$ under the ENN error model. The blue circles represent the tensor $\mathcal{F}_{x_{u(k)}^*,x_{u'(k)}^*,z_{p(k)}^*,z_{p'(k)}^*,f_{ijk}^{x1}, f_{ijk}^{x2},f_{ijk}^{z1}, f_{ijk}^{z2}}(r_k')$. Each green diamond represents an auxiliary tensor $T_{\Lambda^*}$.}
    \label{fig:ez_tn}
\end{figure}

By applying the above construction to the ENN error model, we obtain the tensor network shown in Fig.\ref{fig:ez_tn} for the ENN error model with $d=3$. For non-boundary tensors, each auxiliary tensor $T_{\Lambda^*}$ has dimension $8\times 8\times 8\times 8$. The explicit construction is given in Appendix.\ref{app:ENN}.

\section{performance of Maximum Likelihood Decoding by using multi-index tensor networks\label{Sec:performance}}

The preceding section constructs an efficiently contractible tensor network through two transformations and three steps: reparametrization, elimination, and index classification. Given these crucial results, we can now construct the tensor network for the ENN error model   similar to that of the independent single-qubit error in Ref.\cite{tn_1},  as shown in Fig.\ref{fig:ez_tn}. This means based on our results in prior section, we can directly apply the contraction method of Ref.\cite{tn_1} in our numerical simulations.

In the following numerical simulations, we refer to the maximum-likelihood decoder that computes the coset probabilities using our multi-index tensor network constructed in Sec.\ref{Sec:TN} as the multi-index tensor network decoder. We define $\chi$ as the bond dimension during the contraction. 

Table.\ref{table:p_d3} shows the numerical results of $p(\mathcal{C}(r'))$ in Eq.\ref{eq_pok} for $d=3$, where the exact value is obtained by directly summing over possibilities of all the error configurations. The results obtained with finite bond dimensions agree well with the exact values. We also give the numerical results of $p(\mathcal{C}(r'))$ for $d=15$, as shown in Table.\ref{table:p_d15}.  In this case, direct summation over all error configurations is computationally infeasible, so we compare the results obtained with different bond dimensions. As discussed in Sec.\ref{Maximum Likelihood Decoder}, the maximum likelihood decoder selects the coset with the highest probability. The maximum values among the four coset probabilities shown 
in Table.\ref{table:p_d15} are already stable around $\chi=12$. This indicates that a small bond dimension provides a sufficiently accurate approximation for the tensor network contraction under the ENN error model.

\begin{table}[htbp]
    \centering
    \renewcommand{\arraystretch}{1.2}
    \begin{tabular}{|c|c|c|}
    \hline
     & $10^3\cdot p(\mathcal{C}(rL_I))$ 
     & $10^3\cdot p(\mathcal{C}(rL_X))$ \\ 
     \hline
     exact value 
     & $7.85238486$ 
     & $7.83017389$ \\ 
     \hline
     $\chi = 8$ 
     & $7.85238486$ 
     & $7.83017389$ \\ 
     \hline
     $\chi = 12$ 
     & $7.85238486$ 
     & $7.83017389$ \\ 
     \hline
     $\chi = 16$ 
     & $7.85238486$ 
     & $7.83017389$ \\ 
     \hline
     $\chi = 32$ 
     & $7.85238486$ 
     & $7.83017389$ \\ 
     \hline
    \end{tabular}
    \caption{Comparison between the exact coset probability $p(\mathcal{C}(r'))$ in Eq.\ref{eq_pok} and the results obtained with different bond dimensions $\chi$. The entries in the second and third columns are multiplied by $10^3$. We take $p_{1x}=0.325$, $p_{1y}=p_{1z}=0$, $p_{2x}=0.325$, and $p_{2y}=p_{2z}=0$ for ENN error model.  And for these values of $p_{1x}$, $p_{1y}$, $p_{1z}$, $p_{2x}$, $p_{2y}$, $p_{2z}$, we have $p(\mathcal{C}(rL_Y))=p(\mathcal{C}(rL_Z))=0$. The exact value is obtained by directly summing over possibilities of all the error configurations.}
    \label{table:p_d3}
\end{table}

\begin{table*}[htbp]
    \centering
    \renewcommand{\arraystretch}{1.2}
    \begin{tabular}{|c|c|c|c|c|}
    \hline
     & $p(\mathcal{C}(rL_I))$ 
     & $p(\mathcal{C}(rL_X))$ 
     & $p(\mathcal{C}(rL_Y))$ 
     & $p(\mathcal{C}(rL_Z))$ \\ 
     \hline
     $\chi = 6$ 
     & $8.17944756\times 10^{-76}$ 
     & $1.48128248\times 10^{-73}$ 
     & $2.39746586\times 10^{-70}$ 
     & $3.75290984\times 10^{-74}$ \\ 
     \hline
     $\chi = 8$ 
     & $1.19123905\times 10^{-75}$ 
     & $3.37816888\times 10^{-73}$ 
     & $3.15899214\times 10^{-70}$ 
     & $1.85760861\times 10^{-74}$ \\ 
     \hline
     $\chi = 12$ 
     & $6.43739507\times 10^{-76}$ 
     & $3.94923903\times 10^{-73}$ 
     & $3.31558535\times 10^{-70}$ 
     & $4.02902914\times 10^{-74}$ \\ 
     \hline
     $\chi = 16$ 
     & $6.48830115\times 10^{-76}$ 
     & $3.65136337\times 10^{-73}$ 
     & $3.32193336\times 10^{-70}$ 
     & $3.95768753\times 10^{-74}$ \\ 
     \hline
    \end{tabular}
    \caption{Coset probabilities obtained with different bond dimensions $\chi$ for $d=15$. We take $p_{1x}=0.008$, $p_{1y}=0.008$, $p_{1z}=0.008$, $p_{2x}=0.012$, $p_{2y}=0.008$, and $p_{2z}=0.004$ for the ENN error model. The total error probabilities are $p_1=p_2=0.024$.}
    \label{table:p_d15}
\end{table*}

\begin{figure}
    \centering
    \begin{minipage}{1.0\linewidth}
        \centering
        (a) Multi-index tensor network decoder
        \includegraphics[width=1.0\linewidth]{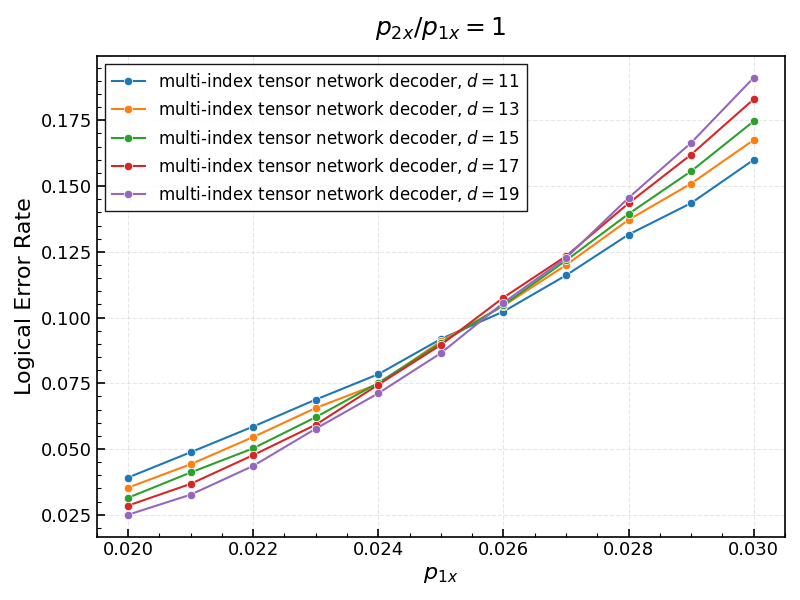}
    \end{minipage}
    \begin{minipage}{1.0\linewidth}
        \centering
        (b) MWPM decoder(PyMatching 2.0)
        \includegraphics[width=1.0\linewidth]{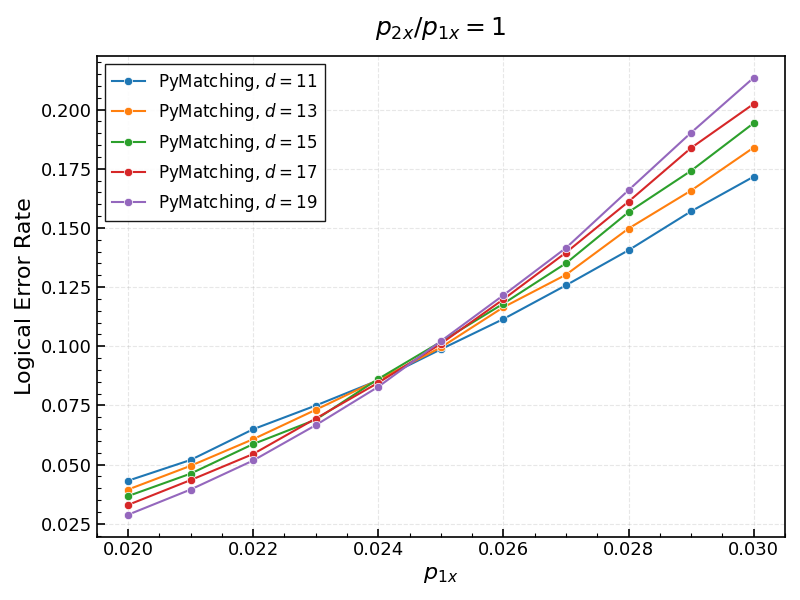}
    \end{minipage}
    \caption{Logical error rate as a function of $p_{1x}$ for the ENN error model with $ p_{1z} = p_{1y} = 0$, $p_{2z} = p_{2y} = 0$ and $p_{1x} = p_{2x}$. (a) Results obtained by our multi-index tensor network decoder. (b) Results obtained by PyMatching 2.0. For our decoder, the tensor network contraction is performed with bond dimension $\chi=12$, and each data point is obtained from 100000 samples.}
    \label{fig:threshold}
\end{figure}

For the ENN error model with $ p_{1z} = p_{1y} = 0$, $p_{2z} = p_{2y} = 0$ and $p_{1x} = p_{2x}$, we estimate the threshold by our multi-index tensor network decoder and compare it with the results from PyMatching 2.0(the MWPM decoder) \cite{MWPM2}. Each data point in Fig.\ref{fig:threshold} is obtained from $100,000$ samples, and the tensor network contraction is performed with bond dimension $\chi=12$. 
Let $p^{cross}_{d_1,d_2}$ denote the crossing point of the logical error rate curves for code distances $d_1$ and $d_2$. We use $p^{th}_d$ to denote the finite-size threshold estimate for the code distance $d$. We estimate $p^{th}_d$ as
\begin{align}
    p^{th}_d
    =
    \frac{1}{7}
    \big(
    p^{cross}_{d-4,d-2}+p^{cross}_{d-2,d}+p^{cross}_{d,d+2}+p^{cross}_{d+2,d+4}+ \\ \nonumber
    p^{cross}_{d-4,d}+p^{cross}_{d-2,d+2}+p^{cross}_{d,d+4}
    \big).
\end{align}
As shown in Fig.\ref{fig:threshold}, for $d=15$, the threshold estimate obtained by our multi-index tensor network decoder is about 2.57\%, while that obtained by PyMatching 2.0 is about 2.46\%. Our decoder shows a finite-size threshold advantage over the MWPM decoder for ENN error model.

However, the estimated threshold is still lower than the asymptotic value in the limit $d\rightarrow\infty$ reported in Refs.\cite{correlation_3,statistical_4}, which is around $2.88\%$. There is no inconsistency here, because our work estimates the threshold at finite code distances, whereas Refs.\cite{correlation_3,statistical_4} report the threshold in the asymptotic limit $d\rightarrow\infty$.
We attribute this difference to finite-size effects, which suggests that the finite-size threshold increases with $d$. In practical implementations, however, surface codes can only have finite code distances. The thresholds shown in Fig.\ref{fig:threshold} are mainly determined by the crossings around $d=15$.

\begin{figure}
    \centering
    \includegraphics[width=1.0\linewidth]{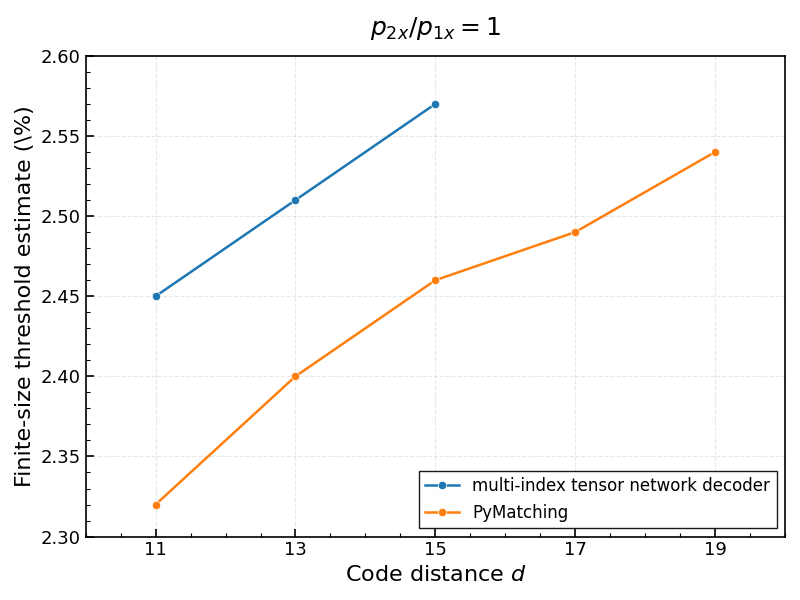}
    \caption{ Estimated finite-size threshold as a function of the code distance
    $d$ for the ENN error model with $ p_{1z} = p_{1y} = 0$, $p_{2z} = p_{2y} = 0$ and $p_{1x} = p_{2x}$. The thresholds obtained by our multi-index tensor network decoder and by PyMatching 2.0 are shown for comparison. Both decoders exhibit an upward trend in the estimated finite-size threshold as $d$ increases.}
    \label{fig:d_threshold}
\end{figure}

The relation between the estimated finite-size threshold and the code distance $d$ for both PyMatching 2.0 and our multi-index tensor network decoder is shown in Fig.\ref{fig:d_threshold}. For both decoders, the estimated finite-size threshold increases with $d$. This upward trend suggests that the threshold estimate under the ENN error model is strongly affected by finite-size effects.

\section{Conclusion\label{Sec:conclusion}}

In this work, we present an explicit tensor network construction for maximum likelihood decoding in the presence of both independent and correlated errors. By using the two transformations introduced in Sec.\ref{Sec:global_TN}, we first construct a tensor network for the noise channel $\mathcal{E}$. However,  since the resulting tensor network contains a huge-dimensional auxiliary tensor, it cannot be contracted efficiently on a classical computer. To overcome this major difficulty, we reconstruct the tensor network through three steps: reparametrization, elimination, and index classification, as described in Sec.\ref{Sec:CTN}. This procedure decomposes the original huge-dimensional auxiliary tensor into a set of lower-dimensional tensors. We also derive explicit conditions that a correlated error model must satisfy for the resulting tensor network to be efficiently contractible on a classical computer. As an example, we apply the construction to the ENN error model.

We have also performed detailed numerical simulations for the ENN error model. The results show that our multi-index tensor network decoder achieves a higher finite-size threshold than the MWPM decoder. 

This shows that our tensor network construction can effectively decompose huge-dimensional auxiliary tensors and thereby produce an efficiently contractible tensor network for correlated error models
satisfying the tractability conditions derived in Sec.\ref{Sec:CTN}. This makes the tensor network method applicable to correlated error models and enables it to achieve improved decoding performance.

\begin{acknowledgements}
We acknowledge the financial support in part by National Natural Science Foundation of China grant No.12174215 and No.12374473, and Innovation Program for Quantum Science and Technology No.2021ZD0300705. This study is also supported by the Taishan Scholars Program.
\end{acknowledgements}

\appendix

\section{Explicit Tensor Network Construction for the ENN Error Model}
\label{app:ENN}

This appendix gives the detailed tensor network construction for the ENN error model. For the ENN error model, each nearest-neighbor data qubit pair is independently affected by a correlated two-qubit noise channel of the form in Eq.\ref{e2}. Let $\ell_{k,j}=\langle k,j\rangle$ denote the nearest-neighbor data qubit pair consisting of data qubits $k$ and $j$, with $\ell_{k,j}=\ell_{j,k}$. When no confusion arises, we simply write this pair as $\ell$. For each pair $\ell=\langle k,j\rangle$, we define the local two-qubit channel
\begin{align}
    \mathcal{E}_\ell(\rho ) =& (1-p_2)\rho + p_{2x}X_{k}X_{j}\rho X_{k}X_{j} + p_{2y}Y_{k}Y_{j}\rho Y_{k}Y_{j} \nonumber \\
    &+ p_{2z}Z_{k}Z_{j}\rho Z_{k}Z_{j}
    \label{el2}
\end{align}

Following the reparametrization introduced in Sec.\ref{Sec:CTN}, the index $d_{\ell yk}$ associated with the YY error can be decomposed into two binary indices $d_{\ell yk,yx}$ and $d_{\ell yk,yz}$. For the ENN error model, it is convenient to eliminate these two indices by allowing a special assignment of $d_{\ell xk}$ and $d_{\ell zk}$. In the original representation, the indices $d_{\ell xk}$, $d_{\ell yk}$, and $d_{\ell zk}$ satisfy the constraint in Eq.\ref{ijk_condition1}. Here, we relax this restriction for $d_{\ell xk}$ and $d_{\ell zk}$ and allow them to take the value 1 simultaneously. This assignment is not interpreted as selecting both the XX and ZZ correlated errors simultaneously. Rather, it is interpreted as selecting the correlated error $Y_kY_j$ in Eq.\ref{el2}. In other words, this assignment plays the role of setting both decomposed YY indices $d_{\ell yk,yx}$ and $d_{\ell yk,yz}$ to 1. Therefore, $d_{\ell yk,yx}$ and $d_{\ell yk,yz}$ do not need to be kept as independent indices and can be eliminated.

Accordingly, $p^*(O_{\ell k})$ in Eq.\ref{p_oik} is modified as:
\begin{align}
    p^*(O_{\ell k}) &= (1-d_{\ell xk})(1-d_{\ell zk})\sqrt{1-p_2} + \nonumber \\
    &d_{\ell xk}(1-d_{\ell zk})\sqrt{p_{2x}} + (1-d_{\ell xk})d_{\ell zk}\sqrt{p_{2z}} +\nonumber \\
    &  d_{\ell xk}d_{\ell zk}\sqrt{p_{2y}}
    \label{p_olk}
\end{align}

\begin{figure}
    \centering
    \includegraphics[width=1.0\linewidth]{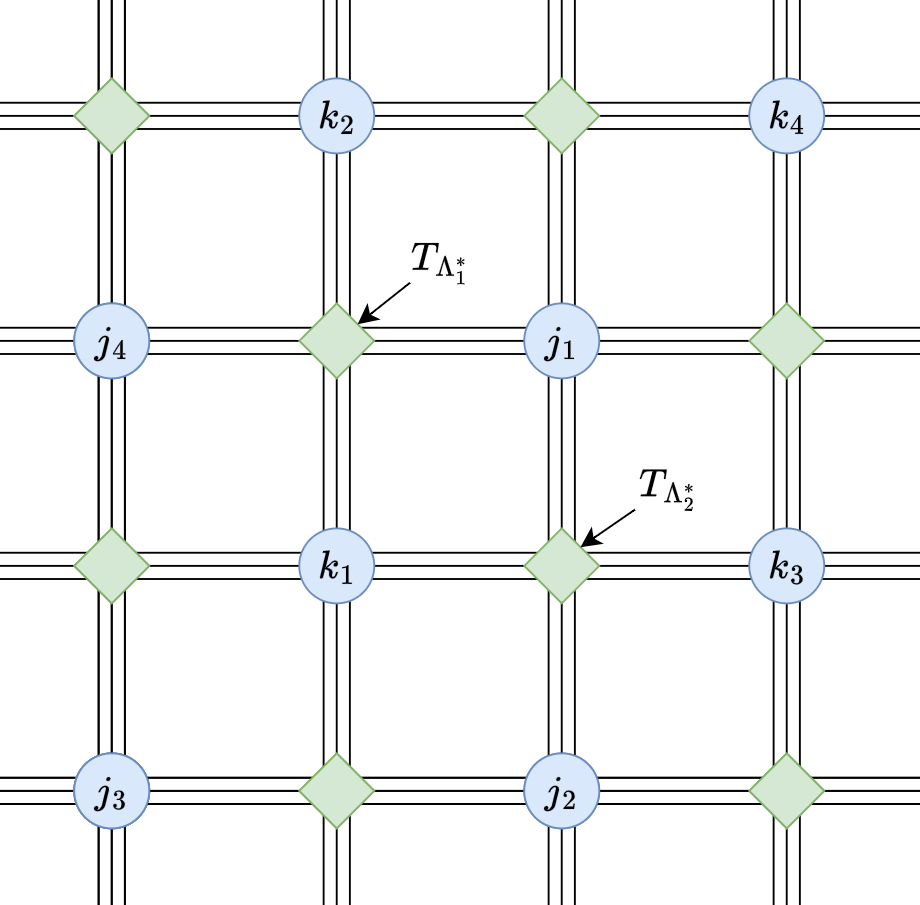}
    \caption{The tensor network corresponding to the surface code under the ENN error model. The blue circles represent the tensor $\mathcal{F}_{x_{u(k)}^*,x_{u'(k)}^*,z_{p(k)}^*,z_{p'(k)}^*,f_{ijk}^{x1}, f_{ijk}^{x2},f_{ijk}^{z1}, f_{ijk}^{z2}}(r_k')$. Each green diamond represents an auxiliary tensor $T_{\Lambda^*}$. The $\mathcal{F}$ tensors are labeled by their corresponding data qubits $k_1,k_2,k_3,k_4$ and $j_1,j_2,j_3,j_4$.}
    \label{fig:tn8_enn}
\end{figure}

After eliminating $d_{\ell yk,yx}$ and $d_{\ell yk,yz}$, we apply the elimination and the classification procedure introduced in Sec.\ref{Sec:CTN}. 
The resulting tensor network for the ENN error model is shown in Fig.\ref{fig:tn8_enn}. In this figure, the $\mathcal{F}$ tensors are labeled by their corresponding data qubits $k_1,k_2,k_3,k_4$ and $j_1,j_2,j_3,j_4$. The explicit forms of the tensors are given below.

We first consider the tensor associated with the data qubit $k_1$.
The sets of indices$\{f_{ijk_1}^{x1}\}$, $\{f_{ijk_1}^{x2}\}$,$\{f_{ijk_1}^{z1}\}$, and $\{f_{ijk_1}^{z2}\}$ introduced in Sec.\ref{Sec:CTN} are given by
\begin{equation}
    \left\{
        \begin{aligned}
        \{f_{ijk_1}^{x1}\}
        &=
        \{d_{\ell_{k_1,j_1}xk_1}, d_{\ell_{k_1,j_4}xk_1}\},
        \\
        \{f_{ijk_1}^{x2}\}
        &=
        \{d_{\ell_{k_1,j_2}xk_1}, d_{\ell_{k_1,j_3}xk_1}\},
        \\
        \{f_{ijk_1}^{z1}\}
        &=
        \{d_{\ell_{k_1,j_1}zk_1}, d_{\ell_{k_1,j_2}zk_1}\},
        \\
        \{f_{ijk_1}^{z2}\}
        &=
        \{d_{\ell_{k_1,j_3}zk_1}, d_{\ell_{k_1,j_4}zk_1}\}.
        \end{aligned}
    \right.
    \label{eq_fijk_enn}
\end{equation}
Using Eq.\ref{p_o1k} and Eq.\ref{p_olk}, we have
\begin{align}
    &\mathcal{F}_{x_{u(k_1)}^*,x_{u'(k_1)}^*,
    z_{p(k_1)}^*,z_{p'(k_1)}^*,
    f_{ijk_1}^{x1},f_{ijk_1}^{x2},
    f_{ijk_1}^{z1},f_{ijk_1}^{z2}}(r_{k_1}')
    \nonumber \\
    &\quad =
    p^*(O_{1k_1})
    p^*(O_{\ell_{k_1,j_1}k_1})
    p^*(O_{\ell_{k_1,j_2}k_1})
    \nonumber \\
    &\qquad \times
    p^*(O_{\ell_{k_1,j_3}k_1})
    p^*(O_{\ell_{k_1,j_4}k_1}) .
    \label{eq:F_k1_enn}
\end{align}

\begin{figure}
    \centering
    \includegraphics[width=1.0\linewidth]{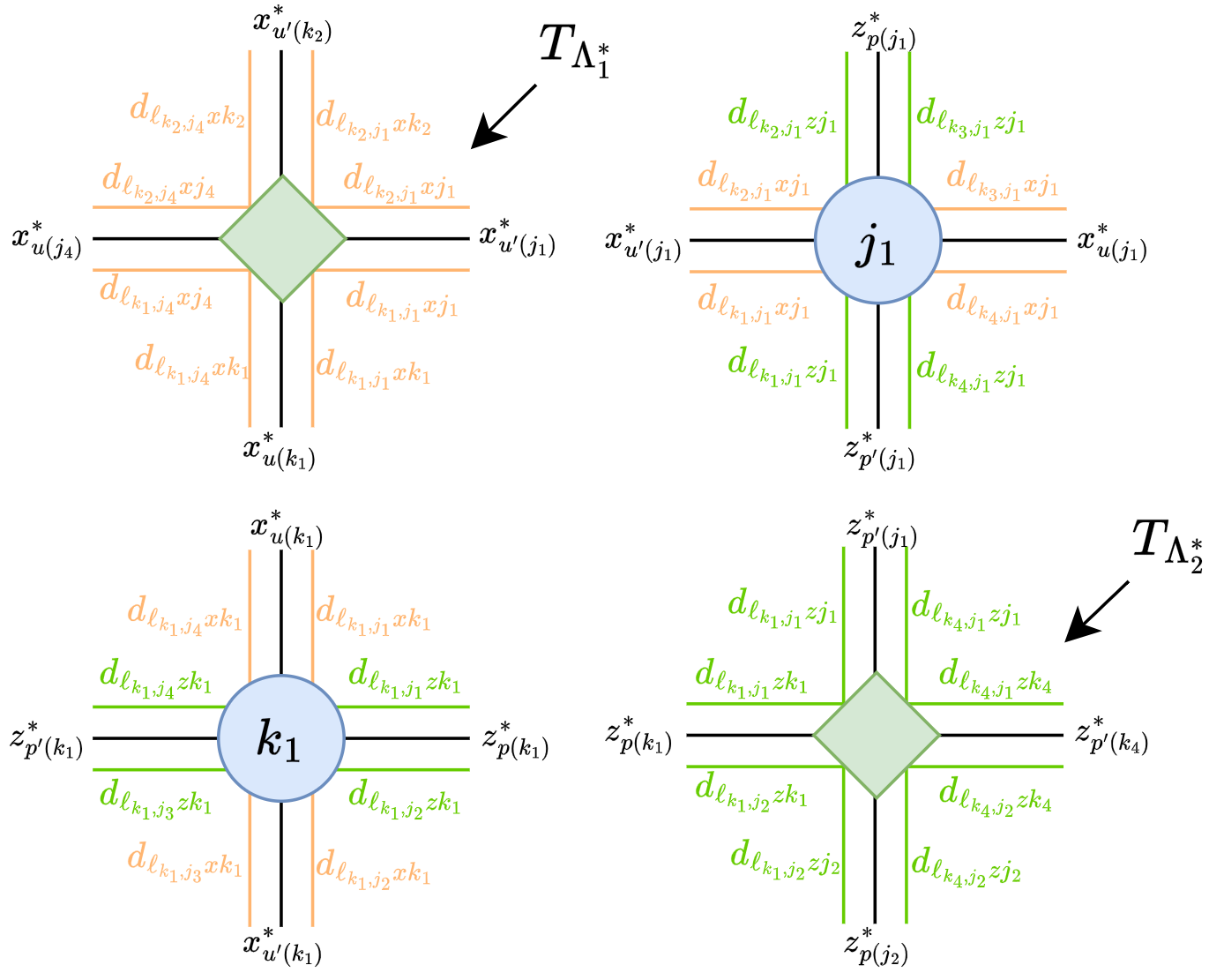}
    \caption{Explicit local forms of the $\mathcal{F}$ tensors and auxiliary tensors $T_{\Lambda^*}$ in the tensor network shown in Fig.~\ref{fig:tn8_enn}.}
    \label{fig:ez_t}
\end{figure}

Next, we give the explicit form of a representative auxiliary
tensor $T_{\Lambda_1^*}$. The auxiliary tensor $T_{\Lambda_1^*}$ is illustrated in Fig.\ref{fig:ez_t}. The corresponding conditions are

\begin{align}
    &x_{u(k_1)}^* \oplus d_{\ell_{k_1,j_1}xk_1} \oplus d_{\ell_{k_1,j_4}xk_1} =  \nonumber \\
    &x_{u'(k_2)}^* \oplus d_{\ell_{k_2,j_1}xk_2} \oplus d_{\ell_{k_2,j_4}xk_2} =  \nonumber \\
    &x_{u'(j_1)}^* \oplus d_{\ell_{k_1,j_1}xj_1} \oplus d_{\ell_{k_2,j_1}xj_1} =  \nonumber \\
    &x_{u(j_4)}^* \oplus d_{\ell_{k_1,j_4}xj_4} \oplus d_{\ell_{k_2,j_4}xj_4}
    \label{app_c0}
\end{align}

and
\begin{equation}
    \left\{
        \begin{aligned}
            d_{\ell_{k_1,j_1}xk_1}
            &=
            d_{\ell_{k_1,j_1}xj_1},
            \\
            d_{\ell_{k_1,j_4}xk_1}
            &=
            d_{\ell_{k_1,j_4}xj_4},
            \\
            d_{\ell_{k_2,j_1}xk_2}
            &=
            d_{\ell_{k_2,j_1}xj_1},
            \\
            d_{\ell_{k_2,j_4}xk_2}
            &=
            d_{\ell_{k_2,j_4}xj_4}.
        \end{aligned}
    \right.
    \label{app_c1}
\end{equation}

The tensor $T_{\Lambda_1^*}$ takes the value 1 when the binary indices satisfy both Eq.\ref{app_c0} and Eq.\ref{app_c1}, and takes the value 0 otherwise.

\newpage

\bibliographystyle{unsrt}
\bibliography{ref.bib}
\end{document}